%% file: sample-sigconf-authordraft.tex
 \documentclass[sigconf]{acmart}
\AtBeginDocument{%
  }

\copyrightyear{2026}
\acmYear{2026}
\setcopyright{cc}
\setcctype{by}
\acmConference[DIS '26]{Designing Interactive Systems Conference}{June 13--17, 2026}{Singapore, Singapore}
\acmBooktitle{Designing Interactive Systems Conference (DIS '26), June 13--17, 2026, Singapore, Singapore}
\acmDOI{10.1145/3800645.3813049}
\acmISBN{979-8-4007-2563-0/2026/06}

\usepackage{graphicx}
\usepackage{soul} 
\usepackage{color} 
\usepackage{pifont}
\usepackage[dvipsnames]{xcolor}
\usepackage{multirow}
\input{sections/commend}
\begin{document}

\title{ConSearcher: Supporting Conversational Information Seeking in Online Communities with Member Personas}
\author{Shiwei Wu}
\email{wushw28@mail2.sysu.edu.cn}
\orcid{0009-0004-3929-455X}
\affiliation{%
  \institution{School of Artificial Intelligence\\Sun Yat-sen University}
  \city{Zhuhai}
  \country{China}
}

\author{Xinyue Chen}
\email{e1520349@u.nus.edu}
\affiliation{%
  \institution{College of Design and Engineering\\National University of Singapore}
  \city{Singapore}
  \country{Singapore}
}
\authornote{Work performed while at Sun Yat-sen University.}

\author{Yuheng Liu}
\email{liuyh357@mail2.sysu.edu.cn}
\affiliation{%
  \institution{School of Artificial Intelligence\\Sun Yat-sen University}
  \city{Zhuhai}
  \country{China}
}

\author{Xingbo Wang}
\email{wangxbzb@gmail.com}
\affiliation{%
  \institution{Bosch Research North America}
  \city{Sunnyvale}
  \state{CA}
  \country{USA}}
\authornote{Work performed while at Weill Cornell Medicine.}
\affiliation{%
  \institution{Weill Cornell Medicine}
  \city{New York}
  \state{NY}
  \country{USA}}

\author{Qingyu Guo}
\email{qguoag@connect.ust.hk}
\affiliation{%
  \institution{Hong Kong University of Science and Technology}
  \city{Hong Kong}
  \country{China}
}
\authornotemark[1]

\author{Longfei Chen}
\email{chenlf@shanghaitech.edu.cn}
\affiliation{%
  \institution{ShanghaiTech University}
  \city{Shanghai}
  \country{China}}

\author{Chuhan Shi}
\email{chuhanshi@seu.edu.cn}
\affiliation{%
  \institution{Southeast University}
  \city{Nanjing}
  \country{China}
}

\author{Zhenhui Peng}
\authornote{Corresponding author.}
\email{pengzhh29@mail.sysu.edu.cn}
\orcid{0000-0002-5700-3136}
\affiliation{%
  \institution{School of Artificial Intelligence \\
Key Laboratory of Intelligent Assessment Technology for Sustainable Tourism, Ministry of Culture and Tourism \\
Sun Yat-sen University}
  \city{Zhuhai}
  \country{China}
}

\renewcommand{\shortauthors}{Shiwei Wu, Xinyue Chen, Yuheng Liu, Xingbo Wang, Qingyu Guo, Longfei Chen, Chuhan Shi, Zhenhui Peng.}

\begin{abstract}
Many people browse online communities to learn from others' experiences and opinions, \eg for constructing travel plans. 
Conversational search powered by large language models (LLMs) could ease this information-seeking task, but it remains under-investigated within the online community. 
In this paper, we first conducted an exploratory study (N=10) that indicated the helpfulness of a classic conversational search tool and identified room for improvement. 
Then, we proposed \name{}, an LLM-powered tool with dynamically generated member personas based on user queries to facilitate conversational search in the community. 
In \name{}, users can clarify their interests by checking what a simulated member similar to them may ask and get responses from diverse members' perspectives. 
A within-subjects study (N=27) showed that compared to two conversational search baselines, \name{} led to significantly higher information-seeking outcome and user engagement but raised concerns about over-personalization. 
We discuss implications for supporting conversational information seeking in online communities.

\end{abstract}




\maketitle

\input{sections/introduction}
\input{sections/relatedwork}

\input{sections/formative}

\input{sections/system}

\input{sections/userstudy}
\input{sections/results}
\input{sections/discussion}
\input{sections/conclusion}





\bibliographystyle{ACM-Reference-Format}
\bibliography{sample-base}
\appendix
\input{sections/ourappendix}

\end{document}

%% file: sections/commend.tex
\newcommand{\eg}{{\it e.g.,\ }}

\newcommand{\ie}{{\it i.e.,\ }}

\newcommand{\name}{{\it ConSearcher}}
\newcommand{\baselineB}{{\it BaseAgent}}
\newcommand{\baselineA}{{\it BaseSearcher}}

\definecolor{mycolor}{RGB}{128,0,128}  

\newcommand{\cxy}[1]{{\color{black} #1}}
\newcommand{\pzh}[1]{{\color{black} #1}}

\newcommand{\revision}[1]{{\color{black} #1}}

\newcommand{{\newrevision}}[1]{{\color{black} #1}}
\newcommand{\majorrevision}[1]{{\color{black} #1}}

\newcommand{\shw}[1]{{\color{black} #1}}
\definecolor{shadow}{RGB}{233, 240, 252}

\newcommand{\chuhan}[1]{{\color{black} #1}}

\newcommand{\wu}[1]{{\color{black} #1}}
\newcommand{\minor}[1]{{\color{black} #1}}

%% file: sections/introduction.tex
\section{Introduction}
Many people come to online communities to seek information from other members who have diverse backgrounds and experiences but have similar interests \citep{liu2011predicting, Kulkarni2013Peer, Oppe2021Hardhats, Arguello2006Talk}. 
For example, people can learn from the member-contributed posts and comments to construct a travel or bodybuilding plan \citep{liu2022planhelper} and acquire visual design knowledge \citep{peng2024designquizzer}. 
\wu{Recently, the integration of LLM-powered conversational search tools into online communities has become increasingly common (\eg Reddit Answer, RedNote's conversational search agent).
While these tools facilitate natural language queries and efficiently summarize relevant information, they tend to flatten the distinct standpoints and rich details embedded in user-generated content which are essential for sensemaking given users' diverse priorities.
For example, when planning a trip, a budget backpacker might prioritize a station’s transport links, whereas a family traveler might find the same station’s crowds overwhelming, preferring a quiet apartment in a residential district instead.
}
\minor{This paper focuses on the emerging paradigm of conversational search within online communities.
}

Previous HCI work has studied the usage of conversational search but has largely focused on its comparisons to traditional web search, \eg regarding biased information querying \citep{sharma2024generative} and user-preferred search tasks
\citep{chi25_conversational_search}. 
It is under-explored how a classic conversational search tool is used and can be improved for information seeking within an online community, which values the personal experiences, opinions, and suggestions contributed by human members \cite{wu2024comviewer,liu2022planhelper}.
\wu{For personalized decision-making tasks (\eg career exploration \cite{malik2024towards}), sensemaking goes beyond fact retrieval. It involves identity construction \cite{weick1995sensemaking} (\ie identify who they are and clarify their latent needs) and dialectical engagement with diverse perspectives \cite{zhang2024see} to understand trade-offs.}
To fill this gap, we first implemented a classic conversational search tool (noted as \textit{BaseAgent}) using community data, which retrieves posts and comments related to the user query, generates an answer grounded on retrieved content, and recommends follow-up questions based on dialog history. 
We then conducted an exploratory study with ten participants to understand their practices and challenges when using \textit{BaseAgent} for information seeking within online communities. 
We found that participants had a back-and-forth interaction between \baselineB{} and original community content, \ie they conversed with \baselineB{} to get compact answers and suggested follow-up queries, and they read posts and comments to situate the information and develop interests in other members' contexts. 
However, participants reported challenges in clarifying personal interests in the queries and getting responses from different members' perspectives,
which limit the helpfulness of \baselineB{}.

To address these challenges,
we introduced the concept of member personas, \ie  
the compact and authentic representations of community members who seek information (noted as \textit{seeker}) of their interests and provide information (noted as \textit{provider}) from different perspectives.  
The seeker personas can help users identify their information needs by looking at members with similar interests in the community, while the provider personas can enable the conversational search tool to offer human-like responses from different members' standpoints.
\wu{HCI researchers have explored the usage of persona to act as target users for eliciting user needs \cite{marsden2017cognitive, eriksson2013secret, csengun2024there, zhu2019creating} and facilitating design decisions \cite{park2022social, proxona,shin2025postermate}, and engage users in thoughtful interaction with social media content \citep{zhang2024see,tanprasert2024debate}.}
\wu{However}, it remains unclear how member personas can be derived and incorporated to support conversational search within an online community.

\wu{To instantiate member personas, we developed \name{},}
which dynamically generates member personas that are grounded on posts and comments related to user queries. 
With \name{}, users can view and customize the generated personas of information seekers, who have clear interests in the factors (\eg travel arrangements, accommodation, and sightseeing) about the general search intent (\eg Japan travel plan). 
Users can select their interested queries that these simulated seekers are likely to ask, and get responses and follow-up queries from the conversational agents role-playing different information providers (\eg anime filmmaker, cultural explorer). 
To enable these system features, we developed an LLM-powered pipeline that 1) decomposes the initial search keywords into related factors, 2) groups posts relevant to each factor for generating personas of information seekers and the queries they are likely to ask, 3) groups comments relevant to these queries for generating personas of information providers, and 4) adopts these personas to guide each agent's response to the user query. 
 

To evaluate the impact of \name{} on \minor{conversational search}
in online communities, we conducted a within-subjects study with 27 participants, asking them to seek information on given topics (\ie make a Japan travel plan, integrate benefits of digital and traditional education, reflect on whether pursuing a PhD or not) in relevant communities.  
The results revealed that compared to \baselineB{} and another baseline (noted as \baselineA{}) with personas of information seekers but not providers, \name{} led to significantly more fruitful 
\minor{learned information}
and engaging experience. 
\name{}'s suggested queries and generated responses were also rated significantly more satisfying, helping users clarify their needs and get contextual answers from diverse perspectives. 
However, the member personas
sometimes complicated the information-seeking process and raised concerns on over-personalization. 

Our contributions are as follows: 
\begin{itemize}
    \item  We conducted an exploratory study to identify the challenges and opportunities for enhancing conversational information seeking within online community contexts. 
    \item We develop \name{}, an LLM-powered system that supports users to clarify information needs and get contextual answers from diverse perspectives by interacting with data-driven member personas. 
    \item Our within-subjects user study demonstrates the strengths of \name{} for improving the outcome and experience of conversational information seeking in online communities.
    \item We provide implications for LLM-powered, persona-driven information-seeking support in online communities.
\end{itemize}


%% file: sections/relatedwork.tex
\section{Related Work}
We first review literature about information seeking support in online communities, which focuses on human members' shared experiences, opinions, and suggestions.
Then, as we propose conversational search and persona methods in online communities, we discuss how each method is utilized in our context. 

\subsection{Information Seeking Support in Online Communities}
\label{sec: information seeking}
Online communities (\eg subreddits in Reddit) are virtual spaces where individuals with shared interests or experiences exchange information, support, and knowledge \citep{waller2019generalists, waller2021quantifying, han2024international, Peng2020Explore, peng2024designquizzer}. 
\wu{For example, youth can learn about professionals' lives and develop career motivation from social media trends like TikTok \#DayInTheLife \cite{malik2024towards}.}
In this paper, we target users who search and read posts and comments within an online community to learn from other members and fulfill their \wu{personalized} information needs. 
Prior HCI researchers have identified user challenges and proposed various intelligent support tools in this information seeking process \cite{zhang2025designing}. 
For example, \citet{wu2024comviewer} found that users had difficulties in finding and making sense of desired content in online mental health communities. 
They proposed a tool that allows users to visually filter community content related to the initial search query via a circle packing view and interactively highlight, summarize, and question any community content \cite{wu2024comviewer}. 
\citet{liu2022planhelper} revealed user challenges in structural information organization and personal need accommodation when learning from question-answering communities for plan construction.
They developed a tool that enables users to read highlighted sentences in answer posts containing key opinions and interactively take notes on the useful points \cite{liu2022planhelper}. 
In general, these tools support users in online communities to search needed information via filters \citep{chen2024amplifying, liu2023coargue, ormel2021using} and digest it via interactive features like note-taking and questioning panels \citep{liu2022planhelper, wu2024comviewer}. 
In line with the goals of these tools but focusing on a different approach, \ie conversational search, our proposed \name{} helps users find and digest their interested community content via conversational agents. 
Our work contributes a novel information seeking support tool in online communities. 

\subsection{LLM-Powered Conversational Search} \label{sec: conversational search}

The rise of large language models (LLMs) boosts the applications of conversational search, 
which empower users to seek and retrieve information through interactive and flexible natural language exchanges \citep{radlinski2017theoretical,vakulenko2019knowledge}, to both open-domain (\eg ChatGPT, Bing Copilot) and domain-specific (\eg documents about certain topics \citep{sharma2024generative}) information seeking tasks.
\wu{To help users navigate through the growing volume of user-generated content, many online communities also integrated conversational search tools in their platform. For example, Reddit provided Reddit Answer \footnote{\url{https://redditinc.com/news/introducing-reddit-answers}}, an AI-powered conversational agent that transforms fragmented information on Reddit into coherent, actionable responses through natural dialogue.}

Recent HCI researchers have compared LLM-powered conversational search to traditional web search in various complex tasks \cite{chi25_conversational_search, sharma2024generative, yang2026fit}.
For example, \citet{sharma2024generative} implemented an LLM-powered conversational search system with a Retrieval Augmented Generation (RAG) approach and studied its impact on users' information seeking behaviors and attitude changes, using tasks of writing essays for controversial topics (\eg ``Should Sanctuary Cities Receive Federal Funding?'').
\wu{They found that LLM-powered conversational search tools can increase people's selective exposure and opinion polarization and LLMs with opinion bias that reinforces users' view could exacerbate bias.}

Similar to these studies, we focus on complex information seeking tasks \cite{suh2023sensecape} that involve multiple aspects or are about controversial topics, \ie Japan travel plan, digital education, and decision on pursuing Ph.D. (\autoref{sec:task_dataset} below). 
Differently, we investigate the usages and designs of conversational search within the online community -- a place that has not been explored for conversational search but is prevalent for information seeking. 
\minor{
Conversational search is particularly promising for complex information seeking within online communities. First, it enhances efficiency by synthesizing vast volumes of fragmented user-generated content into coherent, actionable summaries \cite{sharma2024generative, radlinski2017theoretical}. Second, natural language allows users to articulate nuanced needs through descriptive queries. Moreover, the iterative, multi-turn nature of these dialogues \cite{zhang2024see} enables users to calibrate their requirements as they encounter diverse perspectives, facilitating navigation of the subjective wisdom embedded in community discussions.}
Following \citet{sharma2024generative}, we used an RAG approach to implement a classic conversational search tool with community data. 
We studied its usage via an exploratory study, proposed designs that improve it, and evaluated the proposed tool via a within-subjects study.

\subsection{Application and Development of Personas}
A persona is a fictional but realistic representation of a target person, created based on research, data, and insights about real people \citep{nielsen2014personas}. 
A persona typically features demographic information (\eg age, occupation), goals, and preferences, and it has been applied to fields such as designs, marketing, video creations, health informatics, and so on \citep{nielsen2014personas,shin2025postermate, peng_chi19,proxona,park2022social, csengun2024there}. 
Personas have been shown effective in eliciting user needs and facilitating design decisions \citep{marsden2017cognitive,eriksson2013secret,proxona,shin2025postermate}. 
They are also widely assigned to conversational agents, \eg XiaoIce who is an 18-year-old affectionate girl \citep{zhou2020design} and ComPeer who is a proactive peer supporter with interests in cryptography, basketball, and video games \citep{uist24_compeer}, which can improve personalization and user engagement of the agents \citep{Ha2024CloChat}. 

Recent HCI researchers have utilized LLMs and data-driven methods to generate personas and applied them in various scenarios. 
For example, to support video creators' sensemaking and ideation based on audiences' comments, \citet{proxona} developed an LLM-powered system that transforms audience comments into multi-dimensional (\eg expertise level, motivation) personas, enabling creators to receive simulated feedback and refine work.
\wu{To provide feedback during the research ideation process, \citet{liu2025personaflow} proposed PersonaFlow that generates diverse LLM-simulated expert personas based on research questions and papers.}
Conversational agents with personas were also developed to facilitate critical thinking on social media content \cite{zhang2024see, tanprasert2024debate}. 
For example, \citet{zhang2024see} prompted LLMs to generate five agents with different personas and perspectives on a controversial topic, which helped users engage with unexpected viewpoints via human-like dialogues. 

Inspired by these studies, we leveraged LLMs and community data to generate member personas for eliciting user needs and providing human-like responses from different members' perspectives.
We proposed an LLM-powered pipeline that dynamically generates personas based on users' initial search keywords and relevant community data.  
Our work contributes to the HCI literature on personas with a new and promising application domain -- conversational search in online communities.
\minor{
While conversational search empowers users to engage in iterative, natural language exchanges to navigate community data, articulating concrete needs through prompts remains a challenge for users with vague intentions or limited domain knowledge \cite{tankelevitch2024metacognitive, wang2024task}. Consequently, although the synthesized nature of conversational search is highly effective at consolidating vast volumes of user-generated content into coherent responses \cite{radlinski2017theoretical}, such synthesized summaries can inadvertently flatten the distinct standpoints and individual experiences essential for complex decision-making \cite{yang2026fit}.
Through a within-subjects user study, we explore how simulated personas can assist users in concretizing their latent needs and provide multi-perspective insights.
}

%% file: sections/formative.tex
\section{Exploratory Study}
\label{sec:explorative_study}

\begin{figure*}[]
  \includegraphics [width=\textwidth]{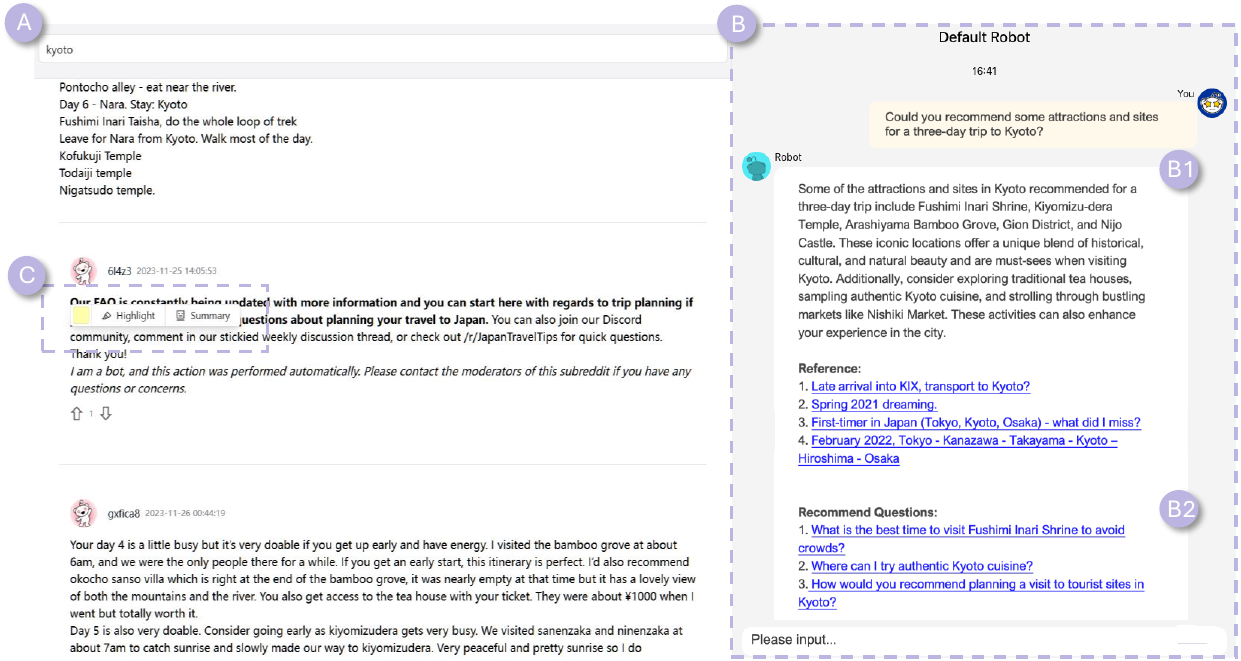}
  \caption{
  \textit{BaseAgent} used in the exploratory study: (A) community interface with posts, comments, and a search box;  (B) \textit{BaseAgent} that B1) answers user queries with links to relevant community content and B2) recommends follow-up questions based on dialog history; (C) pop-up menu that allows users to highlight selected content or prompt \textit{BaseAgent} to summarize it. 
  } 
  \label{fig:baseagent}
\end{figure*}

To explore the usage and design opportunities of conversational information seeking in online communities, we developed \baselineB{} and assessed it with ten participants.

\subsection{BaseAgent in Online Communities} \label{sec:BaseAgent} 
\baselineB{} (\autoref{fig:baseagent}) aims to provide an LLM-powered conversational experience,  similar to 
other conversational search tools (\eg Bing Copilot Search \footnote{\url{https://www.microsoft.com/en-us/edge/features/copilot?form=MA13FJ}},
Reddit Answers \footnote{\url{https://www.reddit.com/answers/}})
in which users converse with an agent using either their own or suggested queries and receive search results as generated texts (often a synthesis).
We chose to implement our \baselineB{} rather than apply the existing industrial-grade ones, which are not open-sourced, are inflexible for customization, and could not directly fit our target scenario, \ie information seeking within one online community. 
For example, Copilot Search targets information seeking using general web resources, while Reddit Answers targets conversational search within all Reddit communities. 
\minor{\baselineB{} allowed us to observe the interaction frictions that arise when the conversational paradigm meets the subjective and experience-based information seeking common in online communities.}

Nevertheless, \baselineB{} shares three key features of these commercial conversational tools and is implemented based on a state-of-the-art approach of Retrieval Augmented Generation (RAG), which is also used in
Copilot Search system architecture \footnote{\url{https://blogs.bing.com/search-quality-insights/february-2023/Building-the-New-Bing}}. 
The three features are answering user queries with relevant content in the community (\autoref{fig:baseagent} B1), recommending follow-up questions based on the dialog history (\autoref{fig:baseagent} B2), and summarizing the content within a page (\autoref{fig:baseagent} A) or any user-selected content (\autoref{fig:baseagent} C). 
We followed \cite{li2024conversational} and used the LangChain \footnote{\url{https://github.com/langchain-ai/langchain}} framework to prompt GPT-4 to split the community posts and comments into meaningful text segments, encoded them with ``all-MiniLM-L6-v2'' model \footnote{\url{https://huggingface.co/sentence-transformers/all-MiniLM-L6-v2}}, and stored them in a Chroma vector database.
Given a query, \baselineB{} first retrieves top-k (k = 5 in our case) relevant texts from the database by cosine similarity. 
We then construct the prompt with both the retrieved documents and the dialogue history.
With prompt engineering, we created prompts to instruct the LLM to generate the responses based on the retrieved texts, recommend three follow-up questions, and summarize the selected text if users request  (See supplementary material for prompt details). 
We used GPT-4 with a 32k context window (gpt-4-32k-0613)
as the backbone model. 
\textit{BaseAgent} is implemented as a web-based application using Python and a Flask backend with React frontend. 


\subsection{Task and Procedure}
\label{sec:task_dataset}

We studied users' practices and challenges when interacting with \baselineB{} for 
\minor{conversational search}
in the online community via an exploratory study with ten participants (four females, six males; age range 19-23, $Mean = 22.30, SD = 1.57$; indexed S01 to S10). 
We recruited them through online advertising and word-of-mouth at a local university.
The inclusion criteria were that they had experience with conversational search (\eg ChatGPT) in general scenarios and information seeking in online communities. 
We randomly assigned each participant to one of three tasks that involve multiple aspects \wu{to consider and are common scenarios for seeking information online}. 

\begin{itemize}
    \item \textbf{Japan travel}: 
    \revision{``Explore the Reddit r/JapanTravel community and create a preliminary Japan travel plan based on your interests''.} 
    This task is similar to those in related HCI work \citep{liu2022planhelper,chi25_conversational_search, chaves2018single} and requires users to form a plan for themselves. 
    \item \textbf{Digital education}: 
    \revision{``Explore the Reddit r/education community and determine how to integrate the benefits of digital and traditional education to suit your own needs''.}
    This task requires participants to form their own opinions on a hot social topic.
    \item \textbf{Ph.D.}: 
    \revision{
    ``Explore the Reddit r/PhD community and reflect on whether pursuing a PhD aligns with your personal goals and circumstances''.
    }
    This task is \wu{about career exploration \cite{malik2024towards} and} closely related to the university students, which is our representative user group, and requires decision-making with gathered information.
\end{itemize}

Accordingly, we collected data on Reddit r/JapanTravel, r/PhD, and r/education from July 10th, 2023, to July 10th, 2024 via Arctic Shift \footnote{\url{https://github.com/ArthurHeitmann/arctic_shift}}. 
We removed the posts and comments whose IDs or content are ``[deleted]'' or ``[removed]'' since we cannot track their relevant comments or which posts they belong to.
After pre-processing, we had $21,220$ / $9,844$ / $32,695$  posts and $449,409$ / $156,826$ / $464,658$ comments for the Japan Travel / Digital Education / Ph.D. task, respectively.

\shw{\textbf{Ethics Disclosure.} We secured the data in firewalled servers, and researchers could download the data only on local machines. Researchers are not allowed to share data and have no interaction with the creators of the collected posts and comments.}

After introducing the task, \textit{BaseAgent}, and corresponding community, we asked participants to spend approximately 30-40 minutes completing the assigned task on a given laptop. 
Throughout the exploration, participants were encouraged to engage in a comprehensive exploration of community content with \baselineB{} and use a document to record key points of their exploration.
After completing the task, participants were asked to share the outcomes of their exploration, describe their search process, and discuss the challenges they encountered with the experimenter. 
Each participant spent around one hour in our study and received 60 CNY as compensation. 
The interviews were recorded, transcribed, and analyzed by two authors using the reflexive thematic analysis method \citep{braun2012thematic}. 
The task process was screen-recorded and replayed by two authors to identify participants' patterns of 
\minor{conversational search}
practices.

\subsection{Findings and Design Goals} \label{sec:practices}
All participants unanimously agreed that \baselineB{} helped them search for information in online communities. 
\shw{``\textit{At first, using the search on the left felt like I was blindly going through content, but the responses on the right helped me connect to some useful information about digital education.}''(S9, F, 22).}
Decomposing the factors related to the search task was their first step. 
Six participants began by browsing posts in the community to gain more information related to the task. 
``\textit{I realized that I should start by thinking about the transportation (of a travel plan) after browsing through different posts.}'' (S3, M, 22).
The other four participants had more clear starting interests and directly queried \textit{BaseAgent}. 
``\textit{I preferred a well-organized trip with minimal travel time between my hotel and attractions, so I asked the chatbot for information about hotels near the sights I wanted to visit.}'' (S5, M, 22). 
During exploration, participants interacted with the community content and \baselineB{} back-and-forth. 
They engaged with \textit{BaseAgent} to get a summary-like answer and follow-up queries to the current query and returned to the sourced posts or comments for contextual information. 
They developed new interests while reading posts and comments and queried \baselineB{}. 
``\textit{After reading a post about someone's trip to Japan, I became interested in the Japanese tea culture. I asked the chatbot for details, and it provided a summary with links to related posts. I clicked a link for further exploration.}'' (S6, F, 23). 
\minor{However, participants reported that while \baselineB{} provided a helpful starting point, they encountered challenges that disrupted the conversational flow. Specifically, the difficulty of articulating vague intentions into specific queries (\autoref{sec:C1}) often resulted in generalized responses that lacked personal relevance (\autoref{sec:C2}). Moreover, the concise and synthesized nature of the agent's output tended to flatten the diverse perspectives expected from a community (\autoref{sec:C3}). These interaction breakdowns often forced participants to abandon the dialogue and revert to manual browsing to regain situational context.}




\subsubsection{Difficulty in thinking of what to query.} \label{sec:C1}
Seven participants (\ie S1, S2, S4, S6, S7, S8, S9) without clear starting interests were a bit unsure what questions they should ask \baselineB{}, especially in the search tasks where they lacked prior experience or knowledge.
``\textit{I could not think of what aspects I could ask \baselineB{} about digital education, as I was vague of its concept.}'' (S8, F, 21). 
Consequently, these participants spent much efforts in filtering and reading original posts at the beginning, making \baselineB{} less helpful in easing this process. 
\shw{
``\textit{Initially, I had no thoughts, but after seeing a post about negative PhD experiences, I realized I should consider the pressure of pursuing a PhD.}''(S2, M, 22).
Participants (\ie S1, S7, S8) also stated that the suggested queries often failed to capture their interest.
S1 (F, 21) mentioned, ``\textit{These questions were superficial and lacked depth. I prefer questions that can inspire me to explore something I am interested in but have not immediately thought of.}'' 
}
Therefore, to improve \baselineB{}, the first design goal (DG) is: 

\shw{\textbf{DG 1}: Help users uncover and express interests of the information-seeking task.}

\subsubsection{\shw{Hard to get responses that match user interests.}}\label{sec:C2}

Seven participants (\ie S1, S2, S3, S5, S8, S9, S10) felt that \textit{BaseAgent} could not understand them well.
S5 (M, 22) stated, ``\textit{The information summarized by the agent did not align with my focuses.}''
\shw{S10 (M, 22) also said, ``\textit{I feel that the concept of digital education is quite broad. As a master's student, the conditions and requirements for utilizing digital education might differ from those of high school students. However, the agent's response was too general and failed to provide information tailored to my specific background and situation.}''}
These issues can be due to that \baselineB{} 
\shw{generates responses that objectively summarize the retrieved community content based on the user query, while also incorporating the context provided by the chat history.
When users fail to clearly express their needs, \baselineB{} cannot provide responses that align with their interests.}
Therefore, we propose:

\shw{\textbf{DG 2}: Help users get responses align with their interests.}



\subsubsection{Challenge in getting responses from diverse perspectives.}\label{sec:C3}
Five participants (\ie S2, S5, S8, S9, S10) agreed that while the agent's responses were concise, they lacked insights from  diverse perspectives, making it difficult for them to think comprehensively about the search goals. 
For example, S10 (M, 22) stated, ``\textit{I hope to understand digital education from the perspectives of different groups, such as high school students, university students, and educators.}''
\wu{S9 (F, 22) stated, ``\textit{I saw a post giving advice specifically for teachers, so I wondered: what about students? So I directly included `for student' in the prompt.}''
}
Similarly, S5 (M, 22) mentioned, ``\textit{I want to design a travel plan suitable for the whole family, including the elderly and children, which requires considering their respective perspectives.}''
\minor{Previous studies indicate that while abstraction aids comprehension efficiency, it often obscures the subjective trade-offs and diverse priorities essential for complex decision-making \cite{keller1987effects, payne1993adaptive}. In the context of online communities, where ``truth'' is often situated in personal experience, a concise summary can hinder a user’s ability to evaluate trade-offs for informed judgment. 
Therefore, for users seeking to learn from shared experiences, it is critical to provide information through diverse member standpoints (\eg the specific concerns of an educator vs. a student) rather than a homogenized overview.
}
To address this challenge, we have:

\textbf{DG 3}: Provide responses from different members' perspectives. 

%% file: sections/system.tex
\section{ConSearcher}
\wu{
Based on the design goals, we introduced the concept of member personas.
In line with the nature of information exchange in online communities, we categorize member personas into information seekers and providers. 
The seekers usually create thread-starting posts to describe their situations and ask for help, 
while the providers leave the comments under the posts to contribute their experiences and insights \cite{peng2021effects,wu2024comviewer}.
In \name{}, member personas serve as exploratory tools that highlight plausible member traits and interests, rather than attempting to replicate real-world community members.
With these member personas, \name{} provides a lens to clarify user interests by looking at the seekers who are similar to them and make sense of the information from different providers' perspectives.
}

\begin{figure*}[]
    \centering
    \includegraphics[width=\textwidth]{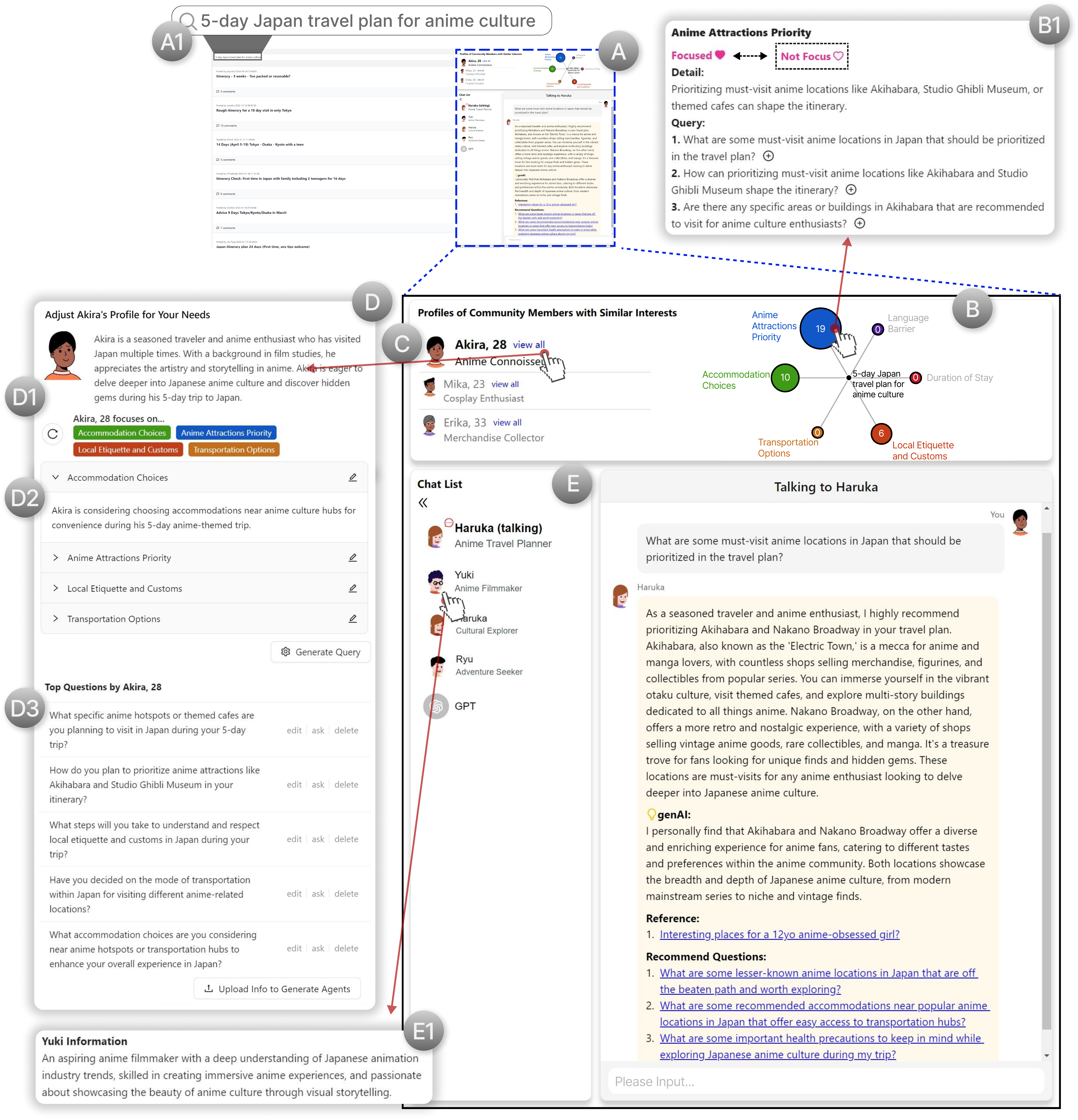}
    \caption{
    \minor{A walkthrough of the \textit{ConSearcher} interface using the 'Japan Travel' task. After an initial query (A1), the system visualizes key dimensions in the Factor Map (B), where circle sizes indicate post volume. Users can then select a Seeker Persona (C) to open a detailed profile (D), which includes the persona’s background (D1), situations for focused factors (D2), and tailored suggested queries generated based on the selected seeker persona (D3) that users can ask to different simulated members in the Provider Persona Panel (E). Each provider persona offers insights based on their specific background and expertise (E1).}
    }
    \label{fig:System}
\end{figure*}

\wu{To construct personas of seekers that inspire search interests, we introduced a factor-situation framework in their profiles. \textbf{Factors} (\autoref{fig:System} B) represent the relevant aspects to the user input query, while \textbf{situations} (\autoref{fig:System} D2) describe the seekers' thoughts on the factors. 
For example, for the factor ``Long-Term Career Goals'' related to the query ``Should I pursue PhD study?'', a possible situation for the seeker persona ``Chris'' is that ``Chris is considering if having a PhD will open up new opportunities in his desired career path or field''.
\minor{The factor-situation framework was developed to bridge the gap between vague intentions and concrete user needs. We introduced \textbf{Factors} to decompose complex search tasks into manageable thematic sub-goals \cite{tankelevitch2024metacognitive, zhang2023examination}. Complementing this, each factor is paired with a \textbf{Situation} to provide detailed scenarios that reflect diverse user preferences.}
Introducing such a framework concretizes the search interests of seeker personas, 
which allows user to clarify and express their needs \minor{by seeing how different goals are articulated in specific contexts} (DG1),
and help the agent generate response that align with user interests (DG2). As for personas of information providers, we kept their profiles brief, including only name, gender, age, identity, and background. 
This design choice, which does not specify factors and situations nor allow user edits, grounds the generation of provider personas that could reflect diverse background of community members and the response only on community data (DG3).}

\subsection{Interface}
The \name{} interface consists of a traditional post-comment reading panel and the conversational search tool in the sidebar (\autoref{fig:System} A). 
To demonstrate how \name{} works, we present a user scenario featuring Harry, a university student who plans a trip to Japan. 
He turns to the Reddit r/JapanTravel community equipped with \name{}. 

\subsubsection{Factor Map (\autoref{fig:System} B): Identifying Interested Factors Related to Initial Query}
The user, Harry, first inputs an initial query ``5-day Japan travel plan for anime culture'' in the search box (\autoref{fig:System} A1) as normally does in online communities.  
\name{} wakes up, and Harry can then view a factor map (\autoref{fig:System} B) visualizing the factors related to the initial query. 
The number within and the size of a node give an overview of how many posts are relevant to that factor and are referenced in the agents' responses as the discussion progresses. 
Harry can left-click each node to filter those factor-related posts in the reading panel and check \chuhan{the details and related queries of} the factor by hovering over a node (\autoref{fig:System} B1).
\wu{For example, when hovering over the ``Anime Attractions Priority'' factor node, Harry sees that the factor means: ``Prioritizing must-visit anime locations like Akihabara, the Studio Ghibli Museum, or themed cafes can shape the itinerary.'' While interacting with agents, he can also ask suggested queries by clicking the ``+'' icon next to each one.}
Harry is interested in this factor and clicks the ``Focused'' icon to indicate his interests, which also adds this factor to seeker persona, as described below. 

\subsubsection{Seeker Persona (\autoref{fig:System} C\&D): Specifying Interests to Get Tailored Suggested Queries} 
\label{sec:seeker_persona}
Next to the factor map, Harry can view three brief profiles of simulated community members dynamically generated based on his initial query and related community posts. 
He is interested in the first one, Akira, who is a 28-year-old identified as an anime connoisseur (\autoref{fig:System} C). He clicks ``view all'' to check its
\wu{background (\autoref{fig:System} D1), focused factors and situations on the factors (\autoref{fig:System} D2)} in a pop-up window (\autoref{fig:System} D). 
\wu{While reading the situations, Harry realizes that he also values accommodation choices but he prefers staying in downtown hotels, so he edits the situations of the focused factors to match his preferences.}
After customizing \wu{persona's profile to elicit his needs and priority,}
Harry clicks the ``Generate Query'' button to get suggested queries tailored to the selected persona, \eg ``What specific anime hotspots or themed cafes are you planning to visit in Japan during your 5-day trip?'' (\autoref{fig:System} D3). 
He can edit or delete a suggested query or ask this query to the agents with different provider personas, as described below which are dynamically generated based on the community comments related to these queries of selected seeker persona. 

\subsubsection{Provider Persona (\autoref{fig:System} E): Getting Contextualized Answers from Agents with Diverse Background} 
Apart from the \baselineB{} (named as GPT in the interface), Harry now has options to converse with agents with different personas of information providers in the Chat List. 
He is interested in learning from the perspectives of Yuki (\autoref{fig:System} E1), who is ``an aspiring anime filmmaker with a deep understanding of Japanese animation industry trends, skilled in creating immersive anime experiences, and passionate about showcasing the beauty of anime culture through visual storytelling''. 
Harry can select suggested queries from the factor map (\autoref{fig:System} B1) or seeker personas (\autoref{fig:System} D3) and ask the agent Yuki, whose answer is structured into \minor{four parts}. 
\wu{\textbf{Answer of provider persona} is grounded in community posts, comments, and the background contexts of both the selected seeker and provider personas.
For example, when Harry asks for anime location recommendations, Haruka, an anime travel planner, provides several locations along with her opinions and the reasons for her recommendations. \textbf{Answer of genAI}, 
is generated entirely by LLM based on the provider persona. 
Given the potential scarcity of community data suitable for specific provider personas, we prompt LLM to provide additional information support. Leveraging their vast training data, LLMs have the potential to provide valuable insights.}
\textbf{Reference} cites community data that contributes to the answer of 
provider persona.
Finally, the response includes \textbf{Recommended Questions} to suggest potential follow-up queries and facilitate Harry’s further exploration.

\begin{figure*}[]
\centering
  \includegraphics [width=0.9\textwidth]{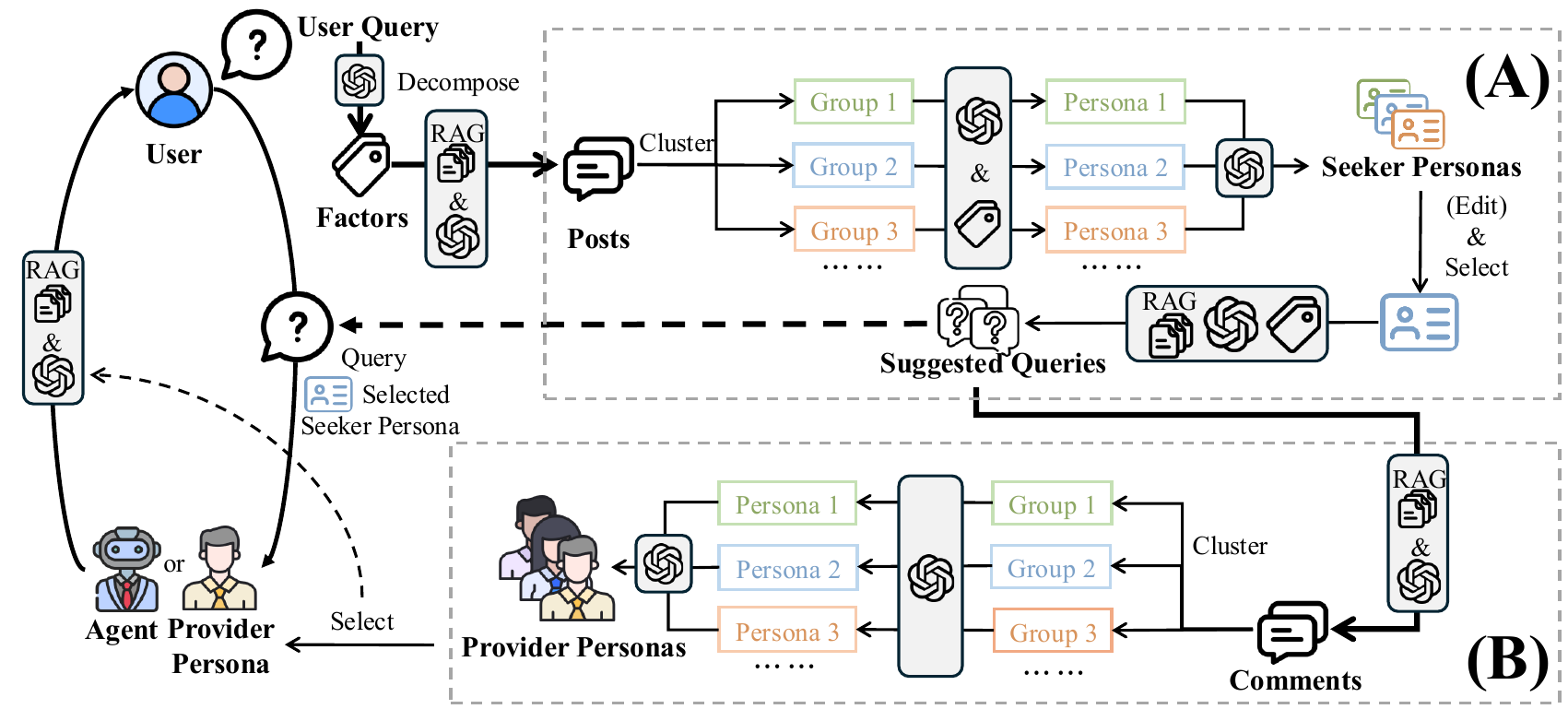}
  \caption{
  Computational workflow of \textit{ConSearcher}'s persona-driven conversational search.
  In the user study, \textit{BaseAgent} does not have the modules of (A) seeker personas and (B) provider personas, while the \textit{BaseSearcher} does not have the module of (B) provider personas.
  }
  \label{fig:personaframework}
\end{figure*}
\subsection{Technical Pipeline}
In this section, we present our technical pipeline (\autoref{fig:personaframework}), detailing the process for factor decomposition, persona generation, and response generation. The pipeline was developed using
OpenAI's gpt-4-1106-preview 
model and demonstrated in the community dataset used in the exploratory study (\autoref{sec:task_dataset}). The prompts used for the pipeline are described in the supplementary materials. 
The parameters in each step are chosen by considering processing time and performance after trials-and-errors. 

\subsubsection{Decomposing Factors Related to the Initial Query}
Given a query in the search box, \name{} first prompts the LLM to decompose the query into several factors, each with a title and an explanation. 
Then, \name{} retrieves top-5 relevant posts to each factor from the community dataset and generates suggested queries (\autoref{fig:System} B1) using the RAG technique.
Besides, when \name{} retrieves posts for response generation as described in \autoref{sec:response_generation}, it prompts LLM to infer what factors these posts discuss and adjusts the nodes' sizes on the factor map accordingly.

\subsubsection{Generating Personas of Information Seekers}
To create seeker personas that suggest tailored queries and facilitate response generation, \name{} first groups all top-5 posts of decomposed factors using HDBSCAN algorithm \citep{campello2013density}. 
For each group, the LLM generates a potential seeker persona (\autoref{sec:seeker_persona}) given the five most central posts and the decomposed factors. 
To avoid generating overly similar seeker personas, \name{} further prompts the LLM to merge and refine these personas to make them distinct. 
If users add a focused factor to a persona and click the ``refresh'' icon (\autoref{fig:System} D1), \name{} will generate a situation of that factor based on the persona. 
Next, \name{} prompts the LLM to suggest five queries that the user-selected persona would personally ask, given the persona \wu{details (\ie background, focused factors with situations)}, user query, \wu{and retrieved posts based on the situations.}

\subsubsection{Generating Personas of Information Providers}
To dynamically create provider personas that can respond to user interests from diverse perspectives, \name{} first applies the RAG technique
to retrieve top-200 relevant comments to all top-five queries of the selected seeker persona. 
We group these comments using HDBSCAN and keep the top-10 central comments for each group. 
We follow \cite{zheng2024disciplink} to adjust the comment groups by prompting the LLM to filter out the groups unrelated or unhelpful to the seeker persona's queries and then divide the remaining groups into comment subgroups, each of which should reflect a similar background experience of a community member. 
Next, we prompt the LLM to generate a provider persona for each adjusted comment group and refine the personas to make them different from each other. 

\subsubsection{Generating Responses}
\label{sec:response_generation}
To generate responses that are tailored to the user query and selected personas of seeker and provider, \name{} first uses the RAG technique to retrieve the top-5 relevant texts (including posts and comments) to the query. 
Then, it filters out the retrieved texts that are very likely irrelevant (\ie cosine similarity score lower than 0.2 between vectorized text and persona's background) to the provider persona. 
Given the remained texts, user query, and backgrounds of selected seeker and provider, \name{} prompts the LLM to generate an answer of provider persona \minor{(\ie Answer of provider persona, Answer of genAI)}. 
To facilitate deep and broad information exploration, \name{} also prompts the LLM to recommend three different questions in the response based on the seeker persona and (1) conversation history, current user query; (2) a randomly selected factor of seeker persona, current user query; (3) a seeker persona's factor which is more likely to be selected if the retrieved texts for response generation talk less about this factor.

%% file: sections/userstudy.tex
\section{User Study} \label{sec: user study}
\begin{figure*}[]
    \centering
    \includegraphics[width=0.99\textwidth]{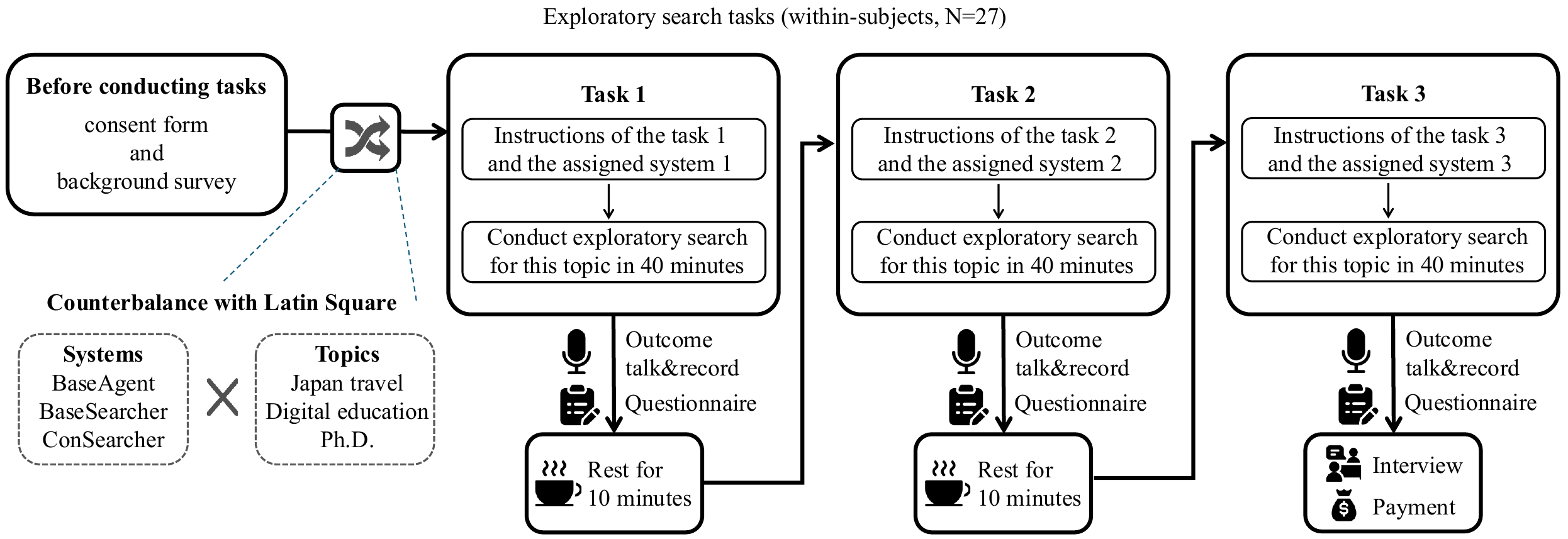}
    \caption{User study procedure. The tasks and systems (\textit{i.e.,} BaseAgent, BaseSearcher, ConSearcher) were counterbalanced using Latin Square, and a post-interview was conducted. The study lasted approximately 3 hours.
    }
    \label{fig:user study}
\end{figure*}
We conducted a within-subjects study with 27 participants to evaluate how \name{}, which is unique with member personas against \baselineB{}, supports conversational search in online communities. 
This study aimed to answer two research questions:

\begin{itemize}
    \item \minor{\textbf{RQ1:} How do member personas (\textit{i.e.,} seeker personas and provider personas) influence the outcomes and processes of conversational search in online communities?}
    \item \minor{\textbf{RQ2:} How do users perceive the utility of \textit{ConSearcher} in supporting conversational search in online communities?}
\end{itemize}

\subsection{Participants}
We recruited 27 participants (15 females, 11 males, one preferred not to disclose; age range 19-29,
$Mean = 21.26, SD = 1.77$, 
indexed P01-P27) through online advertising and word-of-mouth at a local university. 
All participants are not native English speakers but have passed the national College English Test (Level 6) in China. 
They all had experiences in using conversational search tools like Bing Copilot (5 participants daily, 8 4-6 times per week, 14 at least once a week) and seeking information in online communities (12 daily, 5 4-6 times per week, 10 at least once a week). 
The sample size satisfies the G*Power's recommended minimum number (\ie 20) of running ANOVA to compare three conditions in our within-subjects design, given effect size = 0.30 (medium effect), $\alpha$ err prob = 0.05 (default), power (1 - $\beta$ err prob) = 0.8 (an acceptable threshold),
correlation among repeated measures = 0.5 (default), and nonsphericity correction $\epsilon$ = 1 (default).

\subsection{Baselines}
To answer our RQs, we compared \name{} against two baseline conversational search tools. 
All tools share the same community dataset (\autoref{sec:task_dataset}) and have similar interface designs, except for whether having features attached to member personas of information seekers and providers. 

\begin{itemize}
    \item \textbf{\textit{BaseAgent}} is an enhanced version of that in the exploratory study (\autoref{fig:baseagent}) with an additional factor map (\autoref{fig:System} B). It does not have member personas (\autoref{fig:System} CDE). 
    Comparing it to \name{} helps to examine the values of member personas as a whole. 
    \item \textbf{\textit{BaseSearcher}} has the factor map (\autoref{fig:System} B) and personas of information seekers (\autoref{fig:System} CD) but not providers (\autoref{fig:System} E). 
    It generates responses by considering the conversational history, current user query, selected seeker persona, but not provider persona. 
    Comparing it to \baselineB{} and \name{} helps examine the value of seeker persona and provider persona separately. 
\end{itemize}

\subsection{Procedure}

\wu{As shown in \autoref{fig:user study}, after obtaining participants' consent, background, they were required to perform three tasks (\ie Japan travel, digital education, Ph.D.) with assigned tool on the experimenters' computer in a Latin-Square counterbalanced order. Each session began with an introduction of the assigned task and system, followed by conducting task for 30–40 minutes. Afterward, participants verbally recalled learned information from the community about the task and completed questionnaire. There was a 10-minute break between tasks, allowing participants to rest and remain focused. 
In the end, we conducted semi-structured interviews with participants, focusing on their perceptions of the used tools, the generated content, and suggestions for improving \name{} \shw{(\autoref{sec:appI})}. 
Each participant spent approximately 3 hours in our experiment and was compensated 150 CNY (about 21 USD).
}

\subsection{Measures and Analyses}\label{sec:measures} 

\subsubsection{RQ1. Impact on \minor{Conversational Search}}
\minor{Following prior work in community-based information seeking \cite{wu2024comviewer}, we quantified learned information by counting the number of ``meaningful points'' within participants’ transcribed verbal descriptions.} 
We considered a point ``meaningful'' if participants mentioned a meaningful perspective related to the task and integrated it with their own situations. 
For example,
``\textit{The key to online education is that you need to have enough self-discipline, and I believe I can manage that. So, with this self-discipline as a foundation, along with the various benefits of online education, I think digital education suits me quite well.}'' is a meaningful point, 
while ``\textit{First of all, I consider two factors. One is to appreciate the beautiful scenery, the other is the food.}'' is not. 
Two authors first independently coded the meaningful points of 81 verbal descriptions (27 participants $\times$ 3 conditions) and then met and discussed their codes to compile a coding scheme.
They then applied the coding scheme to the rest of the verbal descriptions in a shuffled order and resolved the disagreement through discussions \minor{until full agreement was reached}.

In the questionnaire, we assessed participants' perceived \textit{satisfaction of the received information} (Q1 in \autoref{fig:result}) 
via an item adapted from \citep{liu2022planhelper},
on a standard 7-points Likert Scale (1 - strongly disagree, 7 - strongly agree). 
We assessed \textit{user engagement} in the \minor{conversational search}
process by averaging participants' scores (1 - least engagement, 7 - highest engagement) on six items adapted from Brien's theoretical model \citep{o2016theoretical} (detailed items in \autoref{fig:engagement} in Appendix). 
Besides, we measured their perceived task workload by averaging participants' ratings (1 - least workload, 7 - highest workload) on six metrics adapted from NASA Task Load Index \citep{hart1988development} (\autoref{fig:workload} in Appendix). 

\subsubsection{RQ2. Perception towards Conversational Search Tools}
\label{sec: perception evaluation}
We first measured user perceptions regarding (1) how effectively the system supports users in clarifying their interests while understanding their needs (Q2, Q3 in \autoref{fig:result}), and (2) how well the system delivers responses that facilitate users' sensemaking (Q4, Q5).
We also measured the participants' perceived \textit{helpfulness of the factor map} (applicable to all three tools), \textit{seeker personas} (applicable to \baselineA{} and \name{}), and \textit{provider personas} (applicable to \name{}) for supporting their conversational search in online communities: ``I found the factor map / seeker personas / provider personas helpful in this conversational search task within online communities'' (1 - least helpfulness, 7 - highest helpfulness).  
We asked participants' detailed comments on these 
features during the interview.

To compare the quantitative data among three conditions 
\shw{(\autoref{fig:result})}, we conducted one-way repeated-measures ANOVAs with Bonferroni post-hoc tests. 
For each ANOVA, the assumption of equal variance is confirmed by Mauchly's test of sphericity or otherwise adjusted using Greenhouse-Geisser \citep{girden1992anova}.
We also confirmed via a set of mixed ANOVA analyses that only \shw{2} out of \shw{40} cases have significant main effects of tool/task orders and interaction effects (\shw{detailed results in supplementary material}). 
\shw{
Two authors individually familiarized themselves with the interview content and independently summarized codes into documents.
}
Then, \shw{they}
consolidated themes based on the specific research questions through the thematic analysis process \citep{braun2012thematic}. 
In addition, two authors replayed all the screen recordings of participants' interactions with the assigned tools and counted their frequencies of clicking the references 
to sourced posts, editing the seeker personas, conversing with different provider personas. 

%% file: sections/results.tex
\section{Results}
\label{sec: user study results}

\begin{figure*}[]
    \centering
    \includegraphics[width=0.98\textwidth]{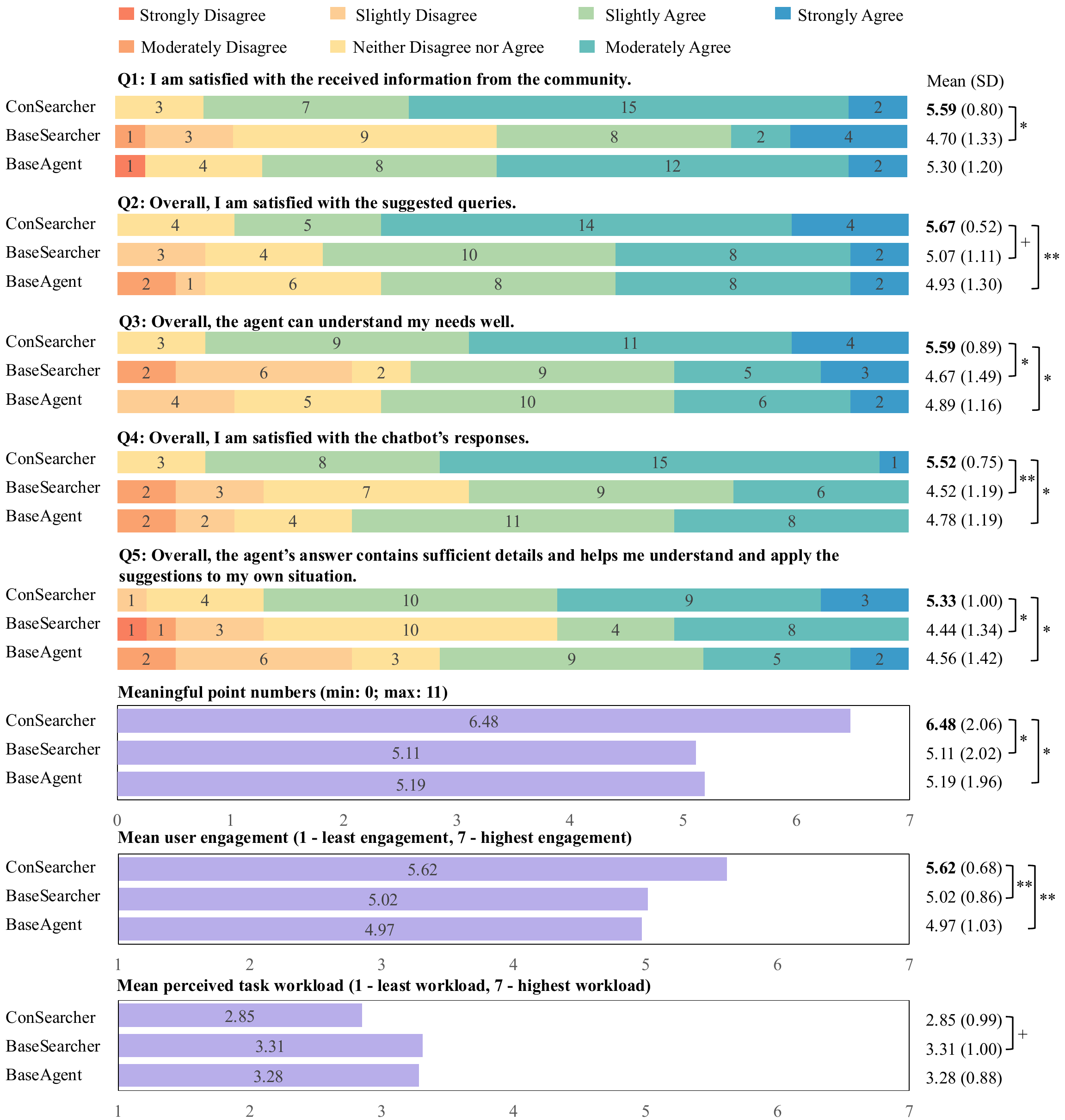}
    \caption{The experiment's statistical results about \textit{BaseAgent}, \textit{BaseSearcher} and \textit{ConSearcher} interface. 
     All items are using one-way repeated measures ANOVA and Bonferroni post-hoc test are then conducted for pairwise comparisons.
     All items except the ``Meaningful point number'' for RQ1 is measured using a standard 7-point Likert scale (1 - strongly disagree/very light burden; 7 - strongly agree/very heavy burden). 
     \majorrevision{The confidence interval is 0.95.} Note:$**:p<0.010;*:p<0.050;+:0.050<p<0.100$ ; within-subjects; N = 27.}
    \label{fig:result}
\end{figure*}

User study results demonstrated that \name{} significantly enhanced participants' \minor{conversational search} outcome in online communities and engagement in this process. 
Participants appreciated the factor map for concreting and tracking the task-related factors, 
seeker personas for clarifying their interests and providing tailored queries,
and provider personas for offering personalized responses from diverse perspectives, 
but reported concerns on complicating and over-personalizing information seeking. 
This section first presents the results of statistical tests and participants' behaviors of conversational search in online communities (RQ1), followed by the findings about participants' perceptions towards the features in \name{} (RQ2). 

\subsection{\minor{RQ1: How do member personas (\textit{i.e.,} seeker personas and provider personas) influence the outcomes and processes of conversational search in online communities?}}
\label{sec:rq1_results}
As shown in 
\shw{\autoref{fig:result}},
there is a significant difference among three conditions regarding the numbers of meaningful points learned from the community; $F(2, 52)=5.876, p = 0.005, \eta^2 = 0.184$.
Specifically, participants with \name{} ($M_{consearcher} = 6.48$) learned significantly more meaningful points compared to the other two conditions ($M_{baseagent} = 5.19, p = 0.013$; $M_{basesearcher} = 5.11, p = 0.012$). 
The perceived satisfaction with the explored information in the community is also significantly different among three conditions; $F(2,52)=5.299, p = 0.008, \eta^2 = 0.169$.
Compared to the \baselineA{} condition, participants with \name{} felt significantly more satisfied with the explored content ($M_{consearcher}=5.59, M_{basesearcher}=4.70, p = 0.019$). 
\minor{There is no significant difference between \baselineB{} condition and \name{} condition ($M_{baseagent}=5.30, p = 0.736$). No significant difference is found between the \baselineB{} condition and \baselineA{} condition ($p = 0.140$).}

As for the mean user engagement in the \minor{conversational search} process, there is a significant difference among three conditions ($F(2,52)=8.611, p <0.001, \eta^2 = 0.249$). 
Participants with \name{} ($M_{consearcher}=5.62$) reported significantly more engaged than when they were with \baselineB{} ($M_{baseagent}=4.97,p=0.007$) and \baselineA{} ($M_{basesearcher}=5.02,p=0.008$). 
Besides, the mean task workload among three conditions has a significant difference ($F(2,52)=3.253, p = 0.047, \eta^2 = 0.169$), though the differences between each two conditions are not significant. 
\minor{This suggests that neither the seeker persona nor the combined persona imposed additional task workload compared to the no-persona baseline. Furthermore, providing both personas did not increase task workload compared to providing only the seeker persona.}

Participants showed different behaviors of interacting with the agents' responses and original community content in three conditions. 
Specifically, participants with \baselineB{} ($M_{baseagent}=5.67$) clicked the source references in the agents' responses significantly more often than when using \name{} ($M_{consearcher}=2.67$, $ p = 0.022$), \minor{whereas their clicking frequency was comparable to that of} \baselineA{} ($M_{basesearcher} = 3.85$, $p=0.832$); $F(2,52)=5.656$, $p = 0.006$, $\eta^2 = 0.179$. 
This suggests that \baselineB{}'s users were more likely to read the original posts and comments, 
while \name{}'s users spent more time acquiring information from the agents. 
Reading original community content took more time but provided valuable contexts for sense-making, 
which could explain \minor{why satisfaction for \baselineB{} remained comparable to that of \name{}, despite the lower information gains.}
\cxy{``\textit{The (\baselineB{}'s) responses were formulaic and general. 
I noticed a referenced post.
The poster advised pursuing a PhD only if deeply passionate about academia, but many commenters disagreed and shared their views. 
This discussion was valuable.}'' (P16, F, 21)}.


\subsection{\minor{RQ2: How do users perceive the utility of \textit{ConSearcher} in supporting conversational search in online communities?}} 
As shown in \autoref{fig:result}, 
participants' satisfaction with \name{}'s ($M_{consearcher} = 5.67$) suggested queries (Q2) was higher than that of \baselineB{} ($M_{baseagent} = 4.93,$ $ p = 0.006$), \minor{while no significant difference was found compared to} \baselineA{} ($M_{basesearcher} = 5.07,$ $ p = 0.053$).
\shw{Participants rated \name{} as the most effective in understanding their needs (Q3: $M_{consearcher} = 5.59, M_{basesearcher} = 4.67, M_{baseagent} = 4.89, F(2,52)=6.734, p = 0.003, \eta^2 = 0.206$).
Participants also rated that \name{}'s agents did the best in providing satisfying and detailed answers that facilitate sense-making
(Q4: $M_{consearcher} = 5.52, M_{basesearcher} = 4.52, M_{baseagent} = 4.78, F(2,52)=5.721, p = 0.006, \eta^2 = 0.180$;
Q5: $M_{consearcher} = 5.33, M_{basesearcher} = 4.44, M_{baseagent} = 4.56,$ $ F(2,52)=4.891, p = 0.011, \eta^2 = 0.158$).
}
We summarize participants' perceived helpfulness of \name{}'s features and corresponding comments as below. 

\subsubsection{Identifying and Tracking Task-Related Factors with the Factor Map}
Participants in all conditions generally rated the factor map helpful ($M_{consearcher}=5.81, M_{basesearcher}=5.70, M_{baseagent}=5.63$) for conversational search. 
For one thing, six participants mentioned that the map helped them break the search task with smaller and more manageable factors at the beginning. 
``\textit{When planning a trip to Japan, the map helped me quickly identify factors like language barriers and budget based on my initial keywords}.'' (P04, F, 20). 
For another, the sizes of factor nodes changed based on related retrieved posts for response generation, helping participants track their progress. 
``\textit{I often checked factors I hadn't explored much and asked myself, `Was I truly not interested in this, or did I just overlook it?'
I also checked factors I had explored more. If I saw a larger node, I knew it was my real interest.}'' (P16, F, 21).

\subsubsection{Benefits of Seeker Personas}
The seeker personas in \name{} ($M_{consearcher} = 5.30$) and \baselineA{} ($M_{basesearcher} = 5.44$) were generally deemed helpful. 
On average, participants in \name{} condition edited the generated seeker personas 1.26 times, while in \baselineA{} they edited the seeker personas 1.70 times ($p = 0.081$). 
In \name{} and \baselineA{},
participants reported two key benefits brought by the seeker personas. 

\textbf{Clarifying Search Interests through Interaction with the Seeker Personas.} 
\minor{Twelve} participants mentioned that reviewing and editing the seeker personas sparked their initial search interests. 
``\textit{I wanted to fly to Osaka but was not sure what was next. The system recommended a reasonable, interesting food enthusiast persona. I modified the age and background to match mine. I then used the food-related suggestions to create a preliminary (Japan travel) plan.}'' (P22, M, 23).
Participants also played roles of multiple seeker personas to learn from their interests, even though the backgrounds of participants and personas were different. 
``\textit{I switched between personas with different backgrounds to clarify my needs. For example, I checked the nature enthusiast's interests in natural landscapes and switched to the culinary expert when I wanted to explore food.}'' (P21, M, 20). 
Interestingly, P01 (F, 23) customized a seeker persona to clarify her interests on a potential future event, 
``\textit{I was currently not interested in pursuing a PhD but was curious how things might change after gaining work experience. I created a persona who is a project manager with 8 years of experience and a background in energy studies. I used this persona to explore whether pursuing a PhD later in my career would be a good choice.}''

\textbf{Getting Queries and Responses Tailored to User Interests.} 
In \name{} and \baselineA{}, the seeker personas were included in the prompts for generating suggested queries (\autoref{fig:System} D3) and responses of the conversational agents. 
\minor{Sixteen} participants mentioned that this helped express their personal circumstances in the queries and get personalized information support from the agents.  
For example, based on \baselineA{}'s suggested queries,  P02 (F, 21) asked \textit{``If I have not worked as a laboratory assistant or in other related jobs, will it be difficult to apply for a PhD?''}. 
P16 (F, 21) reported, \textit{``I customized the (seeker) persona to reflect my interest in anime and Japanese pop culture. The system (\name{}) recommended an anime-themed agent to interact with me, providing travel tips, must-visit spots, and insider insights tailored to my interests.''}

\subsubsection{Strengths of Provider Personas}
Participants found provider personas in \name{} helpful ($M_{consearcher}=5.78$). 
On average, they interacted with 2.30 (ranged 1-5) \name{}'s agents with different provider personas. 
Participants outlined two key strengths of provider personas. 

\textbf{Engaging Users with Human-Like Responses.} 
The significantly improved user engagement in the \minor{conversational search} process with \name{} 
(\autoref{fig:result})
can be attributed to its provider personas, which provided a more immersive and human-like conversational experience \minor{as noted by fifteen participants}. 
P19 (M, 22) shared, ``\textit{I kept chatting with a cultural ambassador to learn from his background and experience.}'' 
P08 (F, 25) reported her experience with an agent who was pursuing a Ph.D.,  \textit{``I eagerly asked him why he chose to do so. He shared the factors he considered in making his decision''}.

\textbf{Providing Responses from Diverse Perspectives.}
Fifteen participants favored engaging with agents with different backgrounds, which provided more comprehensive insights. 
P02 (F, 21) shared her conversations with an elementary school teacher agent about the impact of digital education on students, \textit{``The teacher explained that for young students using iPads, the focus was more on the app usages, which was different from my focus on the underlying computer technology. 
This new perspective helped me gain insights into areas I overlooked.''}
P25 (F, 20) noted, ``\textit{I wanted to see if people with different backgrounds would provide the same viewpoint on a particular question.}''  

\subsubsection{Concerns on Complicating Information Seeking} 
\name{} allows users to edit seeker personas and select provider personas, which helped clarify needs and get contextual responses but sometimes complicated the information-seeking process. 
Specifically, given only a user's initial query in the search box, the generated seeker personas often failed to align with user backgrounds and interests, and editing these personas posed challenges. 
\textit{``I found the seeker personas hard to use. They were largely different from me. I spent a lot of time adjusting a persona but failed, leading to a poor information-seeking experience.''} (P20, M, 20).
Besides, five participants reported that, at times, they struggled to determine which provider persona best aligned with their current query.
P01 (F, 23) noted, ``\textit{I needed to think about the differences between these personas; sometimes, I was unsure which one I should engage with.}''

\subsubsection{Concerns on Over Personalization}\label{sec:overp}
\name{} supports personalization of member personas for conversational information seeking, but participants were concerned that the personas were sometimes over-personalized, reducing their perceived usefulness of and trust on the agents' responses. 
P24 (F, 24) noted, \textit{``Having a persona with a background did provide more detailed responses, but these details were not always useful and were sometimes distracting''}. 
P12 (F, 22) also noted, \textit{``I would like to talk to a person who actually went to Japan before rather than virtual agents, whose information was less trustworthy''}. 
Such user concerns could be partially due to the fact that the over-personalized personas make it less possible to retrieve relevant community content for generating useful answers. 

%% file: sections/discussion.tex
\section{Discussion} \label{sec: discussion}
\minor{While our exploratory study specifically employs \baselineB{} as a probe within the context of conversational search, some findings (\eg the difficulty in clarifying evolving user needs \cite{palani2021conotate, wu2024comviewer}) are also applicable to broader information-seeking behaviors. However, these challenges are amplified in conversational interfaces due to heightened sensitivity to prompt precision \cite{tankelevitch2024metacognitive} and the synthesized nature of responses \cite{sharma2024generative, radlinski2017theoretical}. Unlike traditional search, where users usually browse multiple documents to manually resolve their own vague intentions, conversational search adopts a more 'direct-to-answer' approach. Consequently, the reliance on precise prompting shifts the burden of information filtering from the browsing stage to the query formulation stage, making it critical for future systems to assist users in expressing nuanced needs and providing diverse perspectives to facilitate user sensemaking.}
In this section, we discuss the implications of using data-driven member personas to support conversational search in online communities, use boundaries of \name{}, and the generalizability of our system to support \minor{conversational search} outside one online community. 
Lastly, we briefly discuss the limitations and future work. 

\subsection{Conversational Search in Online Communities with Member Personas: Strengths and Concerns}
In line with previous information seeking support work \citep{liu2022planhelper,liu2023coargue,wu2024comviewer,chen2024amplifying} in online communities, we focus on facilitating users to find and make sense of needed information about human members' experience, opinions, and suggestions within the contexts of posts and comments. 
The exploratory study \minor{(\autoref{sec:explorative_study})} validated the helpfulness of a classic conversational search tool in online communities and identified room for improvements, which were addressed by the introduced member personas. 
Specifically, the seeker personas provided examples of who in the community might have similar interests to the information seeker, helping users clarify their search interests and get tailored queries to and responses from the conversational agents. 
The provider personas were dynamically generated given the retrieved comments that are relevant to the user-selected seeker persona's likely asked queries. 
User study results demonstrated the provider personas helpful for offering comprehensive insights from different people's standpoints. 
In all, these member personas in \name{} led to significantly more effective and engaging conversational information seeking in online communities, compared to the tools with no personas (\baselineB{}) or only seeker personas (\baselineA{}). 

\wu{Our work extends previous applications of personas in HCI literature from providing feedback for ideation \citep{proxona,shin2025postermate, liu2025personaflow, park2022social} and thoughtful critical thinking for user-generated content sensemaking \citep{zhang2024see,tanprasert2024debate} to supporting \minor{conversational search} in online communities.
Leveraging the large volume and diversity of user-generated content in online community and capacity of LLMs to generate high-quality personas \cite{shin2024understanding}, we construct real-time and concrete personas which not only prompts participants to reflect on their own priorities, but also engage with detailed suggestions from diverse perspectives, ultimately enhancing sensemaking for personalized decision-making tasks.}

Despite the demonstrated strengths, the concerns on complicating information seeking and over-personalization require more attentions in future studies. 
To tackle the former concern, 
\minor{future designs could provide pre-defined options for selection and leverage AI to assist in persona generation and updating.
Specifically, instead of requiring manual profile adjustments at the start, the system could present a diverse range of pre-generated situations for rapid selection. 
Furthermore, systems could allow users to create customized personas simply by articulating their core concerns or preferences to the system, which then expands these inputs into a comprehensive profile and offers refined options for confirmation.
As the conversation progresses, the system could then dynamically infer user interests \cite{niu2025part} and update the seeker persona by detecting background cues directly from user queries and interactions.}
\wu{To facilitate provider persona selection, future systems could provide a confidence score calculated based on the user's profile to indicate each provider persona's suitability for the user query.}

The latter concern on over-personalization could cause the lack of community data relevant to the specific queries and user backgrounds, which would amplify the issues of hallucination commonly seen in LLM-powered agents. 
\minor{
Our user study also revealed that \baselineA{} did not outperform \baselineB{} in terms of user satisfaction with the received information.
We attribute this to an expectation mismatch: while seeker personas effectively encourage users to deeply reflect on and concretize their specific information needs, 
the agent’s general responses cannot meet the users' heightened expectations. Consequently, users may mistakenly attribute this poor output to their own failure in defining the seeker persona. This traps them in a frustrating cycle of repeated edits in hopes of yielding better responses—an experience echoed by P20 (\autoref{sec:overp}). This demonstrates that helping users refine their needs without providing corresponding relevant information is ultimately counterproductive. 
}
\name{} also provided answers purely generated by LLMs \minor{(\ie Answer of genAI)} in its response in cases the answers grounded on the community data were not satisfactory.
However, to maximize the values of online communities for information support, the system should help users realize what the community can offer and what it can not to meet their information needs. 
For example, it could visually group the types and amount of information retrieved based on user queries \cite{wu2024comviewer}. 
In cases little existing community data is satisfactory, the system could help users create new posts, \eg by generating topical and emotionally relevant images to the drafted post \cite{cscw25mentalimager}, to express their situations and needs to gain support from human members.
\wu{Moreover, agents could also ask some reflective problems to enhance sensemaking and critical thinking \cite{malik2024towards, peng2024designquizzer} and supplement their response with personalized links that direct to a platform with relevant resources \cite{malik2024towards}.}

\minor{Our findings also highlight a critical challenge in using simulated personas: the conflict between synthetic narratives and authentic human experience. Although \name{} ensures transparency by providing direct links to original community posts, some participants expressed a preference for interacting with real people rather than virtual agents and found overly personalized details distracting (\autoref{sec:overp}). This suggests that in online communities, trust depends heavily on preserving the genuine, unpolished nature of peer-to-peer discourse. When an agent synthesizes community opinions into a flawlessly smooth, heavily personalized narrative, it risks feeling artificial and undermining that trust. 
To maintain trust, future agents should not function solely as synthetic experts that provide efficient, context-aware summaries, they can also act as guides that helps users locate and analyze relevant posts \cite{zhang2024see}. For instance, they can explicitly point out different opinions or debates found in the cited links.}

\subsection{Use Boundaries of ConSearcher}
While \name{} is designed for supporting conversational search in online communities, its features are more or less desirable in certain use cases. 
The seeker personas can be used for \textit{clarifying search interests} when users have vague search goals like ``Japan travel plan'' and can be used for \textit{getting tailored queries} when users struggle to formulate their messages to the conversational agents. 
When users are able to articulate a concrete interest in the query, \eg ``Please recommend some places I can visit in Kyoto. I will be in Kyoto for 3 days and 2 nights.'' (P26, M, 21, in \baselineB{} condition), the seeker personas are less desirable. 
The provider personas facilitate \textit{making sense of sought information from diverse perspectives} when users desire engagement with the community content to achieve their information seeking goals, \eg making a Japan travel plan for themselves, forming their own opinions on digital education, and evaluating whether they should pursue a Ph.D. (\autoref{sec:task_dataset}). 
In cases users would like to gain an overview of community content related to user queries, a conversational search agent that objectively summarizes the retrieved posts and comments point by point 
\footnote{Such summary-like responses are the default option in the recently launched conversational search tools in RedNote community (related news in Chinese: \url{https://news.qq.com/rain/a/20250613A05Z2D00}) and Reddit platform (\url{https://techcrunch.com/2024/12/09/reddit-tests-a-conversational-ai-search-tool/}).} 
could be more desirable. 
Besides, the agents, either with or without provider personas, are undesirable if users attempt to get prediction, \eg ``Is the success rate high?'' asked by P02 about Ph.D. applications in \baselineA{} condition. 

While our studies did not compare the conversational search to the classic search engines embedded in the communities, related HCI work comparing conversational to web search has indicated potential user preferences based on the psychological distance of information seeking tasks \cite{chi25_conversational_search}. 
For example, if the event (\eg a travel) in the task happens in a far future (\eg next year vs. next week) or involves persons far from the information seeker (\eg Internet user vs. close friend) or has not been confirmed (\eg plane tickets not booked vs. booked), users could prefer conversational search over classic web search. 
Future research could explore the user tasks that desire conversational search via deploying \name{}, \eg as a browser extension, within online communities.

\subsection{Generalizability of ConSearcher to Broader Information Seeking Scenarios}
We believe our approach can be generalized to a broader range of information seeking scenarios by including different data sources for persona and response generation. 
Within the community-based settings, we can use our technical pipeline to process the data collected from targeted communities, or all communities within a platform (\eg Reddit) when the data is available, to 
support users to search information about bodybuilding \citep{liu2022planhelper}, design critiques \citep{peng2024designquizzer}, and so on.
For example, the member personas can be incorporated into the conversational search tool Reddit Answer, which would be generated based on the posts and comments in all subreddits that are relevant to the user query. 
In general web-based search or LLM-based search, a deducted version of \name{} without community data can also be applied to enable persona-driven conversational search. 
For example, the seeker and provider personas can be generated based on the search keywords and optionally the data sources in targeted domains, while the interaction design of \name{} can remain for supporting users in clarifying their needs and getting responses from diverse perspectives. 

\subsection{Potential Impact of LLM-Powered Conversational Search on Online Communities}
While this paper targets users who seek information by browsing the community content, we should be aware of the potential impact of the LLM-powered conversational search on these users' contributions back to the community. 
On the positive side, our developed conversational search tools, especially \name{} with member personas, help users effectively acquire needed information from the community, which may enhance their perceived values of the community and willingness to join it. 
On the negative side, when the answer based on member-contributed content is not satisfying or when the LLM could generate more useful answers without community data, users of conversational search tools may reduce their contributions to the community. 
For example, following the release of ChatGPT, posting activity on Stack Overflow decreased by about 16\%, increasing over time to around 25\% within six months \cite{del_Rio_Chanona_2024}. 
Future work could systematically examine how using LLM-powered conversational search tool in online communities would affect users' subsequent behaviors like posting, commenting, and upvoting in the communities. 
\minor{Our user study (\autoref{sec:rq1_results}) also found that participants in \name{} condition clicked significantly fewer source references might reflect higher efficiency, but it could also indicate an over-reliance on the generated content.
If users stop verifying AI responses against the original posts, it may further reduce direct human-to-human interaction. Furthermore, the long-term effectiveness of \name{} depends on a continuous flow of new community content. If community members stop posting, the tool would lose its grounding in authentic, up-to-date experiences, increasing the risk of generating outdated advice or AI hallucinations. 
Conversational search tools should function as cognitive scaffolds that help users navigate complex information, rather than replacing interpersonal interactions.
Therefore, the next challenge for HCI research is to design systems that actively promote user participation rather than just providing answers. Future systems could serve as bridges that encourage users to express their own views. For instance, an agent could detect when a user’s unique situation is not yet covered by existing data and prompt them to share it with the community.}

\subsection{Limitations and Future Work}
While \name{} introduces a novel approach to conversational information seeking support in online communities, we acknowledge several limitations. 

We implemented our conversational search tools that leverage the state-of-the-art RAG technique and support conversational search within one community. 
The values of our proposed member personas may be strengthened if future work is able to implement the personas in an industry-level conversational tool like Reddit Answer that can access all types of communities, which can provide sufficient relevant data for generating personas and responses.
\minor{While this study examined the impact of personas on the conversational search interaction paradigm, the technical evaluation of persona faithfulness—specifically identifying potential hallucinations or biases during synthesis—warrants further investigation. Future work could follow \citet{lei2026humanllm} to evaluate personas through human annotation and model-based assessment, measuring dimensions such as the degree of hallucination.}

We conducted exploratory and user studies mostly with university students, who are representative users but can not represent other user groups of conversational search in online communities. 
\minor{We acknowledged that only involving participants with prior experience in conversational search to minimize the potential confounding effect of novelty could potentially neglect novice users who might exhibit different interaction patterns due to their limited initial exposure to conversational search tools.
Furthermore, although participants passed the CET-6 and articulated their findings in Chinese to facilitate expression, processing English content remains cognitively demanding for non-native speakers. Participants' English proficiency could influence how they formulated queries, comprehended community content, and recalled what they learned. Future research could involve native speakers to further validate the system's effectiveness without the influence of language barriers.
}
The controlled lab studies helped us examine \name{}'s impact on single-time information seeking tasks, but users' interests could evolve during longer-term interactions within the communities. 
Future work needs to test \name{} with more diverse user groups and conducting long-term user studies to evaluate its adaptability and sustained impact in dynamic and evolving contexts.

Our user study largely sought participants' qualitative feedback on how the member personas facilitated conversational search in online communities. 
We did not quantitatively evaluate the accuracy and user trusts of agents' responses, which limits a more comprehensive understanding of their impact on information seeking. Future work could explore additional dimensions to quantitatively evaluate agents' response.
\minor{Counting ``meaningful points'' cannot distinguish between information derived from community-grounded responses and LLM-generated content. Consequently, it is difficult to determine how much each component contributed to the users' learning outcomes. Future work could map each point recalled by participants back to specific interaction logs. This would provide a clearer understanding of the relative contributions of community-grounded responses versus AI knowledge.}

We did not include a condition with provider personas and without seeker personas in the user study, as the generations of provider personas depended on seeker personas in \name{}. 
However, there missed a chance to assessing the independent values of provider personas for conversational search support within online communities. 
Besides, as the user study aimed to evaluate the inclusion of member personas upon a classic conversational search tool, we did not compare \name{} to a baseline condition without conversational search support. 
Similar to the studies that compare conversational search to web search \cite{chi25_conversational_search,sharma2024generative}, future studies should examine the strengths and weaknesses of \name{} compared to the search engines in online communities for various information seeking tasks. 
Our code will be made available for open-source and encourage future researchers to further customize them \footnote{\url{https://github.com/ShionMing/ConSearcher}}.

%% file: sections/conclusion.tex
\section{Conclusion}
In this paper, we proposed and evaluated \name{}, a conversational search tool enhanced by member personas in online communities.  
\name{} decomposes factors related to the initial user query, generates seeker personas that help users to clarify interests and get tailored queries, and generates provider personas that provide human-like responses from diverse perspectives. 
The user study demonstrated that these member personas in \name{} led to significantly improved outcome and experience of conversational information seeking within online communities, shedding light on leveraging data-driven personas to clarify user needs and provide comprehensive answers. 
\begin{acks}
\pzh{
We used AI (in particular large language models) for the following: implementing the functions of conversational search in BaseAgent, BaseSearcher, and ConSearcher, decomposing the factors related to user query, and generating the member personas. 
Details can be found in the relevant sections. 
Authors take responsibility for the output and use of AI in this paper.
}

This work is supported by the General Projects Fund of the Natural Science Foundation of Guangdong Province in China grant (2024A1515012226) and the Natural Science Foundation of Jiangsu Province (Grants No. BK20241300).
We thank Tianqi Song and Haoxiang Fan for their valuable suggestions and support on this work. Finally, we greatly appreciate the reviewers for
their insightful and constructive feedback.
\end{acks}

%% file: sections/ourappendix.tex
\section{Questionnaire Items Used in User Study}
\begin{figure*}[h]
    \centering
    \includegraphics[width=\textwidth]{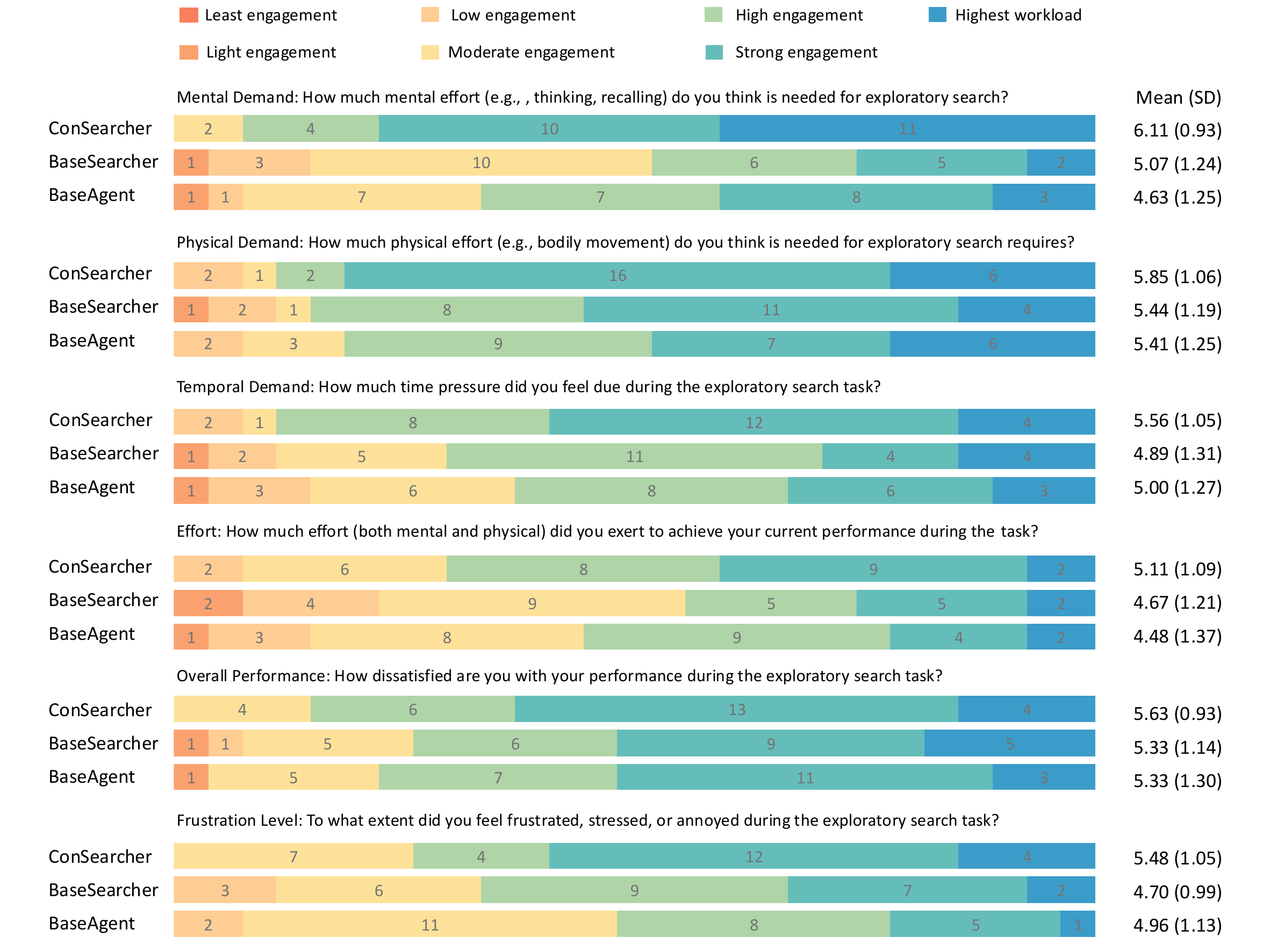}
    \caption{The statistical results about each item on user engagement. 
     All items are using one-way repeated measures ANOVA and Bonferroni post-hoc test are then conducted for pairwise comparisons.
     All items are measured using a standard 7-point Likert scale (1 - least engagement, 7 - highest engagement). 
     \majorrevision{The confidence interval is 0.95.} Note:$**:p<0.010;*:p<0.050;+:0.050<p<0.100$ ; within-subjects; N = 27.}
    \label{fig:engagement}
\end{figure*}

\begin{figure*}[h]
    \centering
    \includegraphics[width=\textwidth]{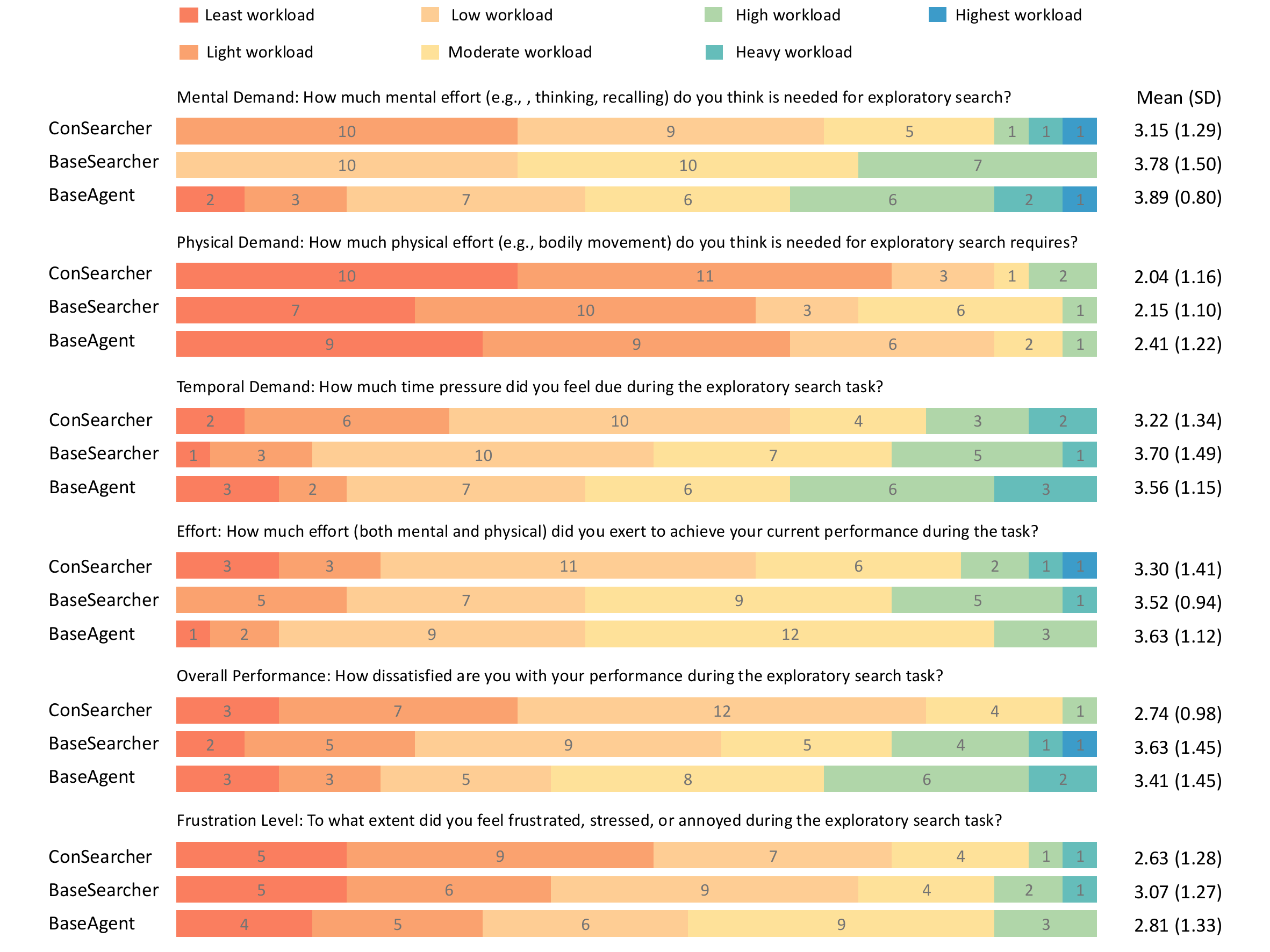}
    \caption{The statistical results about each item on NASA task workload. 
     All items are using one-way repeated measures ANOVA and Bonferroni post-hoc test are then conducted for pairwise comparisons.
     All items are measured using a standard 7-point Likert scale (1 - least workload, 7 - highest workload). 
     \majorrevision{The confidence interval is 0.95.} Note:$**:p<0.010;*:p<0.050;+:0.050<p<0.100$ ; within-subjects; N = 27.}
    \label{fig:workload}
\end{figure*}

\autoref{fig:engagement} and \autoref{fig:workload} provide the detailed items and statistical results of user task workload and engagement.

\section{Interview Questions Used in User Study}
\label{sec:appI}
Please start by discussing your overall experience with this feature (\ie factor map, seeker persona, provider persona), including how it is used, the role it plays during your usage, and whether it has improved the overall information seeking effectiveness and efficiency. How frequently do you use this feature? Please provide specific examples to illustrate your points.
\begin{itemize}
    \item Factor map
    \begin{itemize}
        \item How do you interact with the Factor Map function?
        \item Does the map help you consider factors related to a topic at the beginning?
        \item Do you observe node sizes to balance perspectives and avoid focusing too much on one angle?
        \item Do you click on the nodes to view related posts discussed previously?
        \item Is the assistance provided by the Factor Map worth investing more time and effort in browsing it or clicking on nodes for retrospection?
    \end{itemize}

    \item Seeker persona
    \begin{itemize}
        \item Please briefly describe how you use the Seeker Persona and how you integrate it with the Factor Map.
        \item Does seeker persona help you clarify your needs and priorities? Does it help you think about what to ask, phrase my question.
        \item Do you learn different perspectives from different seeker personas, including those you might not have thought of before?
        \item Is the assistance provided by seeker persona worth investing more time and effort in browsing, selecting, customizing profiles, and refining priorities?
    \end{itemize}

    \item Provider persona
    \begin{itemize}
        \item Please briefly describe and provide examples of how you communicate with different agents.
        \item How do you feel about agents with different backgrounds? Do you think they better suit your needs and understand you more effectively?
        \item How do you feel about the content of the agents' responses?
        \item Among the three recommended questions in the responses, have you noticed differences in their perspectives? During the exploration process, which one do you tend to click on most frequently?
        \begin{itemize}
            \item Do you feel the responses are more personalized and better at understanding your needs?
            \item Do they provide more detailed information?
            \item Do they offer information from multiple perspectives?
            \item Do they make it easier for you to reflect on your own situation?
        \end{itemize}
        \item Is the assistance provided by the provider personas worth investing more time and effort in reviewing the responses, communicating with agents of different backgrounds, or clicking on references to revisit previous interactions?
    \end{itemize}
    \item Additional comments on the feature (\ie factor map, seeker persona, provider persona) and suggestions for improvement.
\end{itemize}